\documentclass[showpacs, prb,twocolumn,preprintnumbers , superscriptaddress, aps]{revtex4-2}
\usepackage[dvips]{graphics}
\usepackage{color}
\definecolor{dred}{rgb}{0,0,0.6}
\usepackage{graphicx}  
\usepackage{dcolumn}   
\usepackage{bm}        
\usepackage{amssymb}   
\usepackage{amsmath}
\usepackage{braket}
\usepackage{upgreek}
\usepackage[mathscr]{euscript}
\usepackage{color}
\usepackage{float}
\usepackage[toc,page]{appendix}
\usepackage[normalem]{ulem}

\usepackage[colorlinks=true, urlcolor=blue, linkcolor=blue, citecolor=blue]{hyperref}

\DeclareUnicodeCharacter{2212}{-}

\begin{document}

\title{Gauge transformation for pulse propagation and time ordered integrals}

\author{Adel Abbout}
\email{adel.abbout@kfupm.edu.sa }
\affiliation{Department of Physics, King Fahd University of Petroleum and Minerals, 31261 Dhahran, Saudi Arabia.}
\affiliation{Advanced Quantum Computing Center,
King Fahd University of Petroleum and Minerals, 31261 Dhahran, Saudi Arabia.}

%%----------------------------------
\begin{abstract}
We investigate a gauge transformation based on the successive elimination of time-dependent onsite potentials at individual sites in finite or infinite systems. Our analysis shows that this transformation renormalizes the inward hoppings by a phase factor $e^{i\phi(t)}$ and the outward hoppings by $e^{-i\phi(t)}$. We further demonstrate how this procedure facilitates the reduction and simulation of pulse propagation in scattering systems, while significantly simplifying the time-ordered integrals involved in the time evolution operator for time-dependent Schrödinger equation.

\end{abstract}

\maketitle

%%----------------------------------------------------------------------------------------------------
\section{introduction}\label{sec1}

%%%
%Time dependent perturbation in mesoscopic systems induces non-equilibrum states that lead to time dependent observables like current, density ad spin pumping. One needs to solve the time dependent Schrodinger equation which usually needs more attention and makes the interpretation of the physical behaviour of the system more challending. Indeed, since energy is not conserved, one needs to deal with infiite sums and integrals correponding to all posible absorbtions/emmission of quantum excitations. In the evolution operator usually involes time-ordered integrals that most of the time need to be numerically. Simplifying the system might lead to simpler expressions and familiar concepts that can be understood in the framework of independnet scattering. 
%In this work, we would like to give introduce a useful gauge transformation for time dependent perturbation (pulses, periodic, raise of potential ...) and show its utility in few examples.
%We are interested in hamiltonians of the form 
%%%

Time-dependent perturbations in mesoscopic systems generate non-equilibrium states that result in observables such as time-dependent currents \cite{Gaury2015}, density modulation, and spin pumping \cite{OLY1,OLY2}. Analyzing these effects typically requires solving the time-dependent Schrödinger equation, which demands particular care \cite{fleury} and often complicates the interpretation of the system’s physical behavior\cite{Cao_2025,heatandcharge}. Since energy is not conserved in the presence of external driving, one must consider infinite sums or integrals corresponding to all possible absorption and emission processes of quantum excitations \cite{Thermo,Kloss_2021}.

Moreover, the evolution operator generally involves time-ordered integrals, which in most cases need to be evaluated numerically \cite{Michael1}. Simplifying the system or choosing an appropriate representation can lead to more transparent expressions and connect the problem to familiar concepts such as independent scattering processes \cite{Stronglybiased}.

In this work, we investigate a convenient gauge transformation for handling time-dependent perturbations (such as pulses, periodic driving, or sudden potential changes)\cite{OLY1,OLY2,fleury,Thermo} and demonstrate its utility through several illustrative examples. We focus on Hamiltonians (Not necessarily Hermitian) of the form:
\begin{equation}
    \mathscr{H}=-\sum_{\left(k, k^{\prime}\right) \in \text { hop }} \gamma^{kk'} c^\dagger_{k} c_{k^{\prime}}-e \sum_{k \in \text { list }} V_k(t) c_k^{\dagger} c_k
    \label{hamiltonian}
\end{equation}
where $c_k^\dagger$ ($c_k$) is the creation (anihiliation) operator of an electron at site $k$ and $\gamma^{kk'}$ is the hopping parameter between two sites (from site $k'$ to site $k$). The parameter "list" refers to the list of sites in the system and "hop" refers to all the possible hoppings between pairs of sites. The time-dependent potential $V_k(t)$ is taken to be site-dependent. $e$ is the electric charge. This Hamiltonian is the tight-binding representation of the Schrödinger operator on a regular lattices but can also  work for more general cases like when defined on a graph (Random Matrix Theory)\cite{Abbout1,Abbout2,Abbout3}. We would like to know what is the effect of eliminating the potential $V_l(t)$ on the specific site of index $l$ using the operator:
\begin{equation}
    U_l\equiv e^{-i\phi_l(t)c_l^\dagger c_l} = \sum_{n=0}^\infty \frac{(-i \phi_l)^n}{n!}\left(c_l^{\dagger} c_l\right)^n
\end{equation}
with the phase $\phi_l(t)$  defined as :
\begin{equation}\label{phase}
\phi_l(t)=\frac{e}{\hbar} \int^t_{-\infty} V_l(t') d t'
\end{equation}
The operator $U_l$ can be simplified if someone recalls that for Fermions  $\left(c_l^{\dagger} c_l\right)^n=c_l^{\dagger} c_l, \forall n \neq 0$ and $c^\dagger_kc^\dagger_k=c_kc_k=0$ We can straightforwardly find that 

$$
U_l=1+z_l c_l^{\dagger} c_l
$$
with $z_l=e^{-i \phi_l(t)}-1$.
The new Hamiltonian with this gauge transformation will be :
\begin{equation}
    \tilde{\mathscr{H}}=U_l^{\dagger} \mathscr{H} U_l-i \hbar U_l^{\dagger} \partial_t U_l \label{transf}
\end{equation}

simple algebra shows that 
%$$U_l^\dagger c_k^\dagger c_{k'} U_l =(1+z_l) c_k^\dagger c_{k'}$$
\begin{equation}
U_l^\dagger c_k^\dagger c_{k'} U_l=
\begin{cases}
(1+z_l^*) c_k^\dagger c_{k'} & l = k\neq k' \\
(1+{z}_l)c_k^\dagger c_{k'} & l = k'\neq k \\
c_k^\dagger c_{k'} &  \text{else} \\
\end{cases}
\end{equation}
Applying this to eq. \ref{transf} the Hamiltonian eq.\ref{hamiltonian} becomes
\begin{equation}\label{equation6}
\begin{split}
\tilde{\mathscr{H}}=&-\sum_{\left(k, k^{\prime}\right) \in \text { hop}'} \gamma^{k k^{\prime}} c_k^{\dagger} c_{k^{\prime}}-e \sum_{k \in \text { list}'} V_k(t) c_k^{\dagger} c_k\\
&-\sum_{k'\in \text{list}'}\gamma^{lk'} e^{+i\phi_l(t)} c_l^\dagger c_{k'}-\sum_{k\in \text { list}'}\gamma^{kl}e^{-i\phi_l(t)} c_k^\dagger c_l
\end{split}
\end{equation}

\begin{figure}
    \centering
    \includegraphics[scale=0.3]{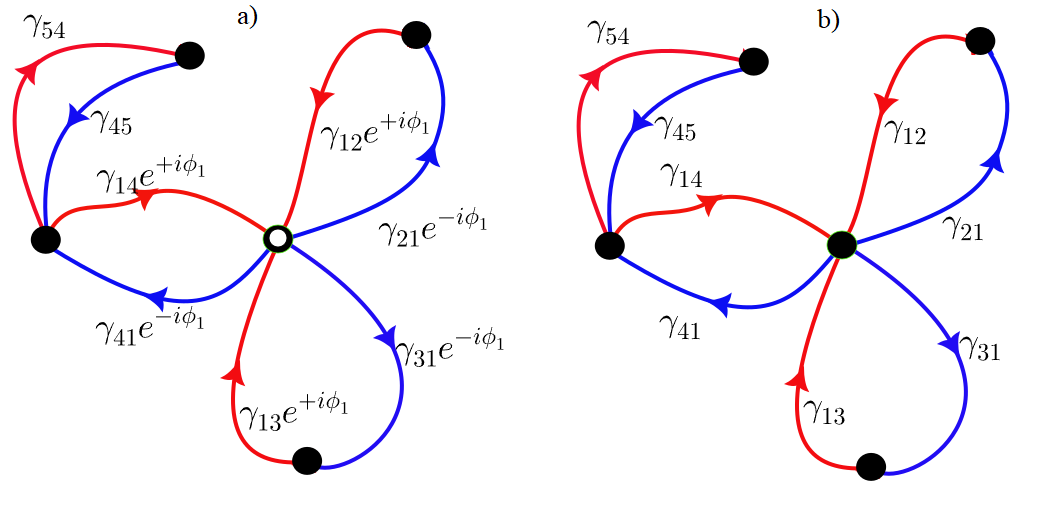}
    \caption{The gauge transformation effect on a graph. The hoppings from the central site (site 1) are renormalized by a phase factor $e^{+i\phi_1}$ when they are inward and by $e^{-i\phi_1}$ when they are outward. It is worth noting that $\gamma_{45}$ and $\gamma_{54}$ are not changed since they do not involve the site 1 }
    \label{fig:gauge}
\end{figure}

where $\text{hop}'$ and $\text{list}'$ are the list of hoppings and sites that do not involve the site $l$  respectively.

To explain Eq. \ref{equation6} with simple words, the elimination of the potential $V_l(t)$ at a given site $l$ requires renormalizing the hopping terms connected to that site. Specifically, the outward hopping from site $l$ acquires a phase factor $e^{-i\phi_l(t)}$, while the inward hopping acquires $e^{+i\phi_l(t)}$, as illustrated in Fig.~\ref{fig:gauge}. All other hoppings that do not involve site $l$ remain unchanged. 

It is worth noting that this gauge transformation does \textit{not} require the Hamiltonian to be Hermitian, and one does not need to eliminate the full potential $V_l(t)$. Instead, one may remove any portion of the potential and adjust the hopping phases accordingly (similarly to Eq.~\ref{phase}). This procedure can be repeated for any number of sites in the system.

\medskip

\begin{figure}
    \centering
    \includegraphics[scale=0.3]{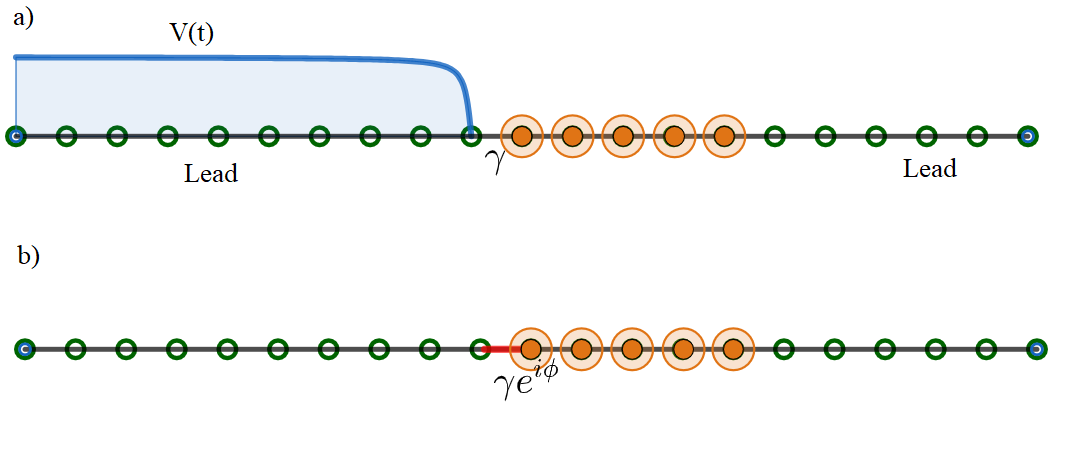}
    \caption{ a) 1D scattering system (orange color) connected to two semi-infinite leads. The left lead is rised to a potential $V(t)$. b) The potential $V(t)$ can be eliminated by a gauge transformation leading to unchanged hoppings in the lead and only the hopping at te interface (in red color) is affected by a phase factor $e^{\pm i\phi(t)}$ (depending on the  hopping left to right or the opposit.).   }
    \label{fig:leadgauge}
\end{figure}

\noindent\textbf{Examples of use:} To study the propagation of an electric pulse in an infinite system, consider a scattering setup composed of a finite central region connected to two semi-infinite leads (see Fig.~\ref{fig:leadgauge}). The potential in one of the leads is raised from zero to a maximum value and then brought back to zero following a time profile $V(t)$, which defines the pulse. The function $V(t)$ may be Lorentzian, Gaussian, triangular, or take any other shape.

Most time-dependent Schrödinger equation solvers (e.g., \texttt{tkwant}~\cite{Kloss_2021}) cannot \textit{directly} handle a time-dependent potential applied to an infinite region. Instead, they perform the gauge transformation described above on all sites of the lead. In this process, each hopping within the lead is renormalized twice, once for each of its two neighboring sites, acquiring opposite phase factors. These phases cancel, leaving all hoppings in the lead unchanged, \emph{except} for the hopping at the interface with the central system.

As shown in Fig.~\ref{fig:leadgauge}, the resulting system therefore contains only a {finite number of time-dependent hoppings}, located at the interface between the lead and the scattering region. This special case where $V(t)$ is uniform in the lead explains why for exmple tkwant puts the phase hoppoing at the interface \cite{Kloss_2021}.\\
\textit{time dependent potential on a finite system:} 
If a uniform potential is applied to a finite system, a gauge transformation removes the time dependence from the central region while transferring it entirely to the hoppings that connect the system to the leads. These interface hoppings acquire the usual phase factor $e^{\pm i \phi(t)}$.

\begin{figure}
    \centering
    \includegraphics[scale=0.4]{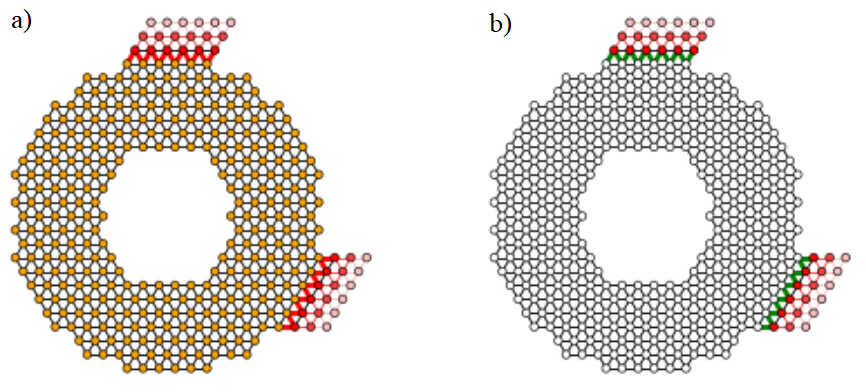}
    \caption{ Eliminating the uniform onsite potential (orange) in (a) results in a central region without any potential (white sites), while the hoppings within the central system remain unchanged. This occurs because neighboring sites contribute opposite phase factors that cancel out. Only the hoppings at the interface are modified, acquiring the corresponding phase factors (the thick red bonds in (a) become thick green in (b))  }
    \label{ring}
\end{figure}

In Fig.~\ref{ring} (a), the ring-shaped system is subject to a uniform time-dependent onsite potential (shown in orange). Applying the gauge transformation site by site eliminates this potential everywhere inside the ring. As a result, all hoppings within the central system remain unchanged: the phase factors associated with neighboring sites cancel exactly on each internal bond. Only the hoppings at the interfaces acquire a phase factor $e^{\pm i \phi(t)}$, as shown in Fig.~\ref{ring} (b).

This example highlights a key point: increasing the potential uniformly in the lead is gauge-equivalent to decreasing it uniformly in the central system. After the transformation, the time dependence survives solely at the interface between the lead and the system.
%If a uniform potential is applied on a finite system, performing the gauge transformation will lead to a central system independent of %time whereas all the outward hopping defining the interfaces with the leads will aquire the usual phase factor $e^{-i\phi(t)}$.In %Fig\ref{ring} a), the system with a ring shape has a uniform time dependent potential (depicted in orange). The gauge transformation %done repeadtedly on each site will lead to a central system, without potential and the hoppings at the interface aquiring  the same %phase factor $e^{i\phi(t)}$ The hoppings inside the system remain unchanged because the phase factors cancel between two neighbooring %sites. This system shows us that increasing the potential (uniformaly) inside the lead is equivalent to lowering it (uniformaly) in the %system. the time dependent will appear only at the interface. 

\textit{time ordered integrals:} the time evolution operator $\mathscr{U}(t)$ for the time-dependent Schrödinger equation is expressed using time ordered integrals,
$$\mathscr{U}(t)=\mathcal{T} \exp \left[-i \int_0^t \mathscr{H}\left(t^{\prime}\right) d t^{\prime}\right]$$

\begin{figure}
    \centering
    \includegraphics[scale=0.35]{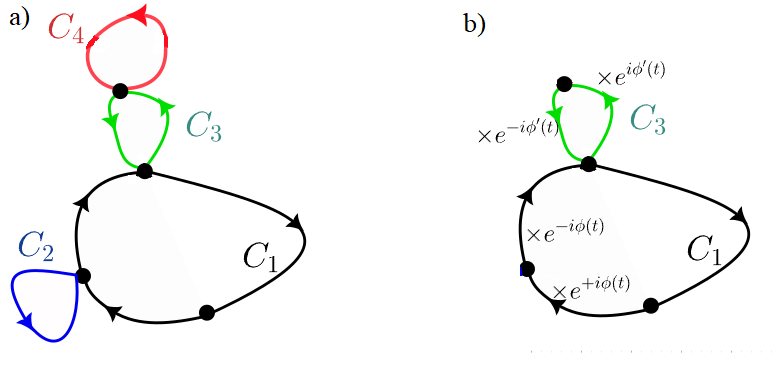}
    \caption{ A graph representing the Hamiltonian matrix. The self-loops (red and blue) represent the onsite potentials. To reduce the number of possible paths starting from a given  node (a site) to itself, one can get rid of the self loops by a gauge transformation leaving only $C_1$ and $C_3$ with renormalized hoppings by a phase factor.  }
    \label{timeorder}
\end{figure}

This integral is in general difficult to  evaluate when the Hamiltonians at two different times do not commute ($[\mathscr{H}\left(t\right) ,\mathscr{H}\left(t^{\prime}\right) ]\neq0$), even for small matrices \cite{Giscard2}. For finite-size Hamiltonians, the matrix representation of $\mathscr{H}\left(t\right)$ can be viewed as a graph, where each site is  connected by inward and/or outward hoppings, and the onsite-potential $V_k(t)$ represented as a self-loop (hopping to itself) as depicted in Fig. \ref{timeorder}. An interesting way to evaluate this time-ordered integral is to use the \text{$\star$}-product (star product) introduced in \cite{Giscard3,Giscard2,Giscard}. The idea is to sum the contribution of all possible \textit{prime} walks (which do not visit any site more than once) including the self-loops. For  clarity, we briefly recall the basis of this calculation using the prime walk formalism, and refer the reader to consult \cite{Giscard,Giscard2,Giscard3} for more details.
A walk that starts from a given site $\alpha$ and returns to it after visiting other sites at most once is considered as a cycle $c_1$. This cycle can be travelled multiple times. The contribution of all these repeated possible loops is $\sum_n c_1^n=1 /\left(1-c_1\right)$. If during the cycle $c_1$ one can bifurcate to a different cycle $c_2$, then one must replace $c_1$ by $c_1\times 1 /\left(1-c_2\right)$. This process can be repeated if another different cycle $c_3$ is in the path defining $c_1$, leading to $c_1\times 1 /\left(1-c_2\right) \times 1 /\left(1-c_3\right)$. Any cycle with  an embedded sub-cycle will thus aquire a multiplicative factor $1/(1-c)$ corresponding to the sub-cycle.
Of particular interest are the self-loops arising from the onsite potentials. For example, the graph in Fig.\ref{timeorder} will yield the following term:
\begin{equation}
\sum_{w \text { walk: } \alpha \rightarrow \alpha} w=\frac{1}{1-c_1 \frac{1}{1-\textcolor{blue}{c_2}} \frac{1}{1-\textcolor{green}{c_3} \frac{1}{1-\textcolor{red}{c_4}}}} .
\end{equation}

In this expression, the self-loops depicted in Fig. \ref{timeorder}, $C_2$ and $C_4$ are onsite potential to which we can apply the gauge transformation detailed previously. This transformation eliminates the self-loops and renormalizes the hoppings as shown in Fig. \ref{timeorder} b). The resulting expression becomes:

\begin{equation}
\sum_{w \text { walk: } \alpha \rightarrow \alpha} w=\frac{1}{1-\tilde{c}_1 \frac{1}{1-\textcolor{green}{\tilde{c}_3} }}.
\end{equation}

$\tilde{c}_3$ and $\tilde{c}_4$ are defined with the renormalized hopping which incorporate the phase factor $e^{\pm i\phi(t)}$. Importantly, what is performed here is not merely a redefinition or  a simple absorbtion of the term $(1-c_2)^{-1}$ in $c_1$. Instead, this transformation effectively circumvents the explicit computation  of terms of the form $(1_\star-f)^{\star-1}=\sum_{n\leq 0} f^{\star n} $  which correspond to the Neumann expansion under the non-commutative $\star$-product.

The non-commutative $\star$-product is defined by the convolution-like relation

$$ (f \star g)(t',t)=\int_t^{t'} f(t',\tau) g(\tau, t) d\tau$$

where the iterated $\star$-powers are recursively constructed as:

$$f^{\star n+1} =f\star f^{\star n}$$
and the $\star$-identity is given by   $1_\star(t',t)=\delta(t'-t))$
The gauge transformation reduces the number of self-loops by eliminating the onsite potentials, thereby simplifying the expression by reducing the number of complex integrals.\\

\textit{precessing spin}: The gauge transformation can be used in the opposit way: we can remove a phase factor from a hopping term and replace it by an onsite potential. To examine such situations, let us consider the case of a single spin precessing at the driving frequency $\omega$. The Hamiltonian is $\mathscr{H}(t)= \vec{\sigma}\cdot\vec{s}(t)$ where the Pauli matrices define $\vec{\sigma}=(\sigma_x,\sigma_y,\sigma_z)$. For a spin  precessing about the $z$-axis, we have (with $s_0^2+s_z^2=1$, where $s_0$ is the inplane component $s_z$ the component along $z$-axis)

$$\mathscr{H}(t)=\begin{pmatrix}
+s_z & s_0e^{-i\omega t} \\
s_0e^{+i\omega t} & -s_z
\end{pmatrix}
$$

by doing the gauge transformation (backward) twice to eliminate the phase factor, we can  prove that:
$$\tilde{\mathscr{H}}(t)=\begin{pmatrix}
+s_z+\omega/2 & s_0 \\
s_0 & -s_z-\omega/2
\end{pmatrix}
$$
(it is easier to test the gauge transformation backward and go from $\tilde{\mathscr{H}}$ to ${\mathscr{H}}$)

The hamiltonian $\tilde{\mathscr{H}}$ is time-indepdendent which makes the calculation easier  for quantities like the spin density $\langle\sigma_z\rangle$.
This procedure shows its effect for larger system where a chain of precessing spins \cite{spiralmagnet,OLY1}(Ferromagnetic material) is adjacent to a normal metal material allowing to express the pumped spin current through the interface using different relations than the spin-mixing conductance \cite{OLY1,OLY2}

\textbf{Conclusion}
Successively eliminating the onsite time-dependent potential can reduce the portion of the Hamiltonian that explicitly depends on time. This, in turn, makes it possible to simulate systems where the time dependence resides primarily in the lead (of effectively infinite size), as encountered in pulse-propagation or spinpumping problems. In such situations, the role of the interface becomes crucial.

Moreover, the complexity of expressing the time-evolution operator-typically complicated by the noncommutative star product-can be significantly reduced, since all self-loops in the graph representing the Hamiltonian can be removed through this procedure. It is also worth noting that the method does not require the Hamiltonian to be Hermitian.
%Successive elimination of the onsite time dependent potential might lead to reduce the size of the part that depends on time in the Hamiltonian and thus allow the %simulation of system primarly with time dependence in the lead (infinite size) for pulse propagation or spin pumpin problems. The role played by the interface in such %situation is crucial. The complexity of expressing the time evolution operator (due to the non-commutative star product) can be reduced since one can eliminate all the %self-loops in the graph representin the Hamiltonian. It is worth noting that we do not need to have a hermitian Hamiltonian to apply this procedure.

\textbf{Aknowledgment}
We acknowledge the support provided by King Fahd University of Petroleum and Minerals (KFUPM) for funding this work through the Interdisciplinary Research Center for Advanced Quantum Computing under grant INSS2512.

\textbf{Funding}
This work has been funded by the interdeciplinary center for advanced quantum computing through the grant INSS2512.
\appendix
\textbf{Data availability}
All data generated or analysed during this study are included in this published article.
\section{calculation}
if $l\ne k'$ then $\{c_{k'},c_l\}=\{c_{k'}^\dagger,c_l^\dagger\}=0$. 
This means that $c_{k'}$ can switch order with two successive operators without additional sign, so 
\begin{align*}
U_l^\dagger c_l^\dagger c_{k'}U_l=& (1+z_l^* c_l^{\dagger} c_l)c_l^\dagger c_{k'}(1+z_l c_l^{\dagger} c_l)\\
=&(1+z_l^* c_l^{\dagger} c_l)c_l^\dagger (1+z_l c_l^{\dagger} c_l)c_{k'}\\
=&(1+z_l^* c_l^{\dagger} c_l)c_l^\dagger c_{k'}\\
=&(1+z_l^* (1-c_l c_l^{\dagger}))c_l^\dagger c_{k'}\\
=&1+z_l^* c_l^\dagger c_{k'}
\end{align*}

where we use the fact that $c_l^\dagger c_l^\dagger=0$ and $c_l^\dagger c_l=1-c_l c_l^{\dagger}$.

In the case where $l\neq k$ and $l\neq k'$ then 

$U_l^{\dagger} c_k^{\dagger} c_{k^{\prime}} U_l= U_l^{\dagger} U_l c_k^{\dagger} c_{k^{\prime}}= c_k^{\dagger} c_{k^{\prime}}$
\bibliographystyle{plain}
\bibliography{references}
\end{document}